\documentclass[reprint,twocolumn,superscriptaddress, amssymb,aps,pra]{revtex4-1}

\usepackage{graphicx}
\usepackage{xcolor} 
\usepackage{amsmath}

\newcommand{\ket}{\right\rangle}
\newcommand{\bra}{\left\langle}

\begin{document}
\title{Molecular Laser-Cooling in a Dynamically Tunable Repulsive Optical Trap}
\author{Yukai Lu}
\affiliation{Department of Physics, Princeton University, Princeton, NJ 08544 USA}
\affiliation{Department of Electrical and Computer Engineering, Princeton University, Princeton, NJ 08544 USA}
\author{Connor M. Holland}
\affiliation{Department of Physics, Princeton University, Princeton, NJ 08544 USA}
\author{Lawrence W. Cheuk}
\affiliation{Department of Physics, Princeton University, Princeton, NJ 08544 USA}

\date{\today}
\begin{abstract}
Recent work with laser-cooled molecules in attractive optical traps has shown that the differential AC Stark shifts arising from the trap light itself can become problematic, limiting collisional shielding efficiencies, rotational coherence times, and laser-cooling temperatures. In this work, we explore trapping and laser-cooling of CaF molecules in a ring-shaped repulsive optical trap. The observed dependences of loss rates on temperature and barrier height show characteristic behavior of repulsive traps and indicate strongly suppressed average AC Stark shifts. Within the trap, we find that $\Lambda$-enhanced gray molasses cooling is effective, producing similar minimum temperatures as those obtained in free space. By combining in-trap laser cooling with dynamical reshaping of the trap, we also present a method that allows highly efficient and rapid transfer from molecular magneto-optical traps into conventional attractive optical traps, which has been an outstanding challenge for experiments to date. Notably, our method could allow nearly lossless transfer over millisecond timescales.

\end{abstract}
\maketitle
Ultracold polar molecules, with their rich structure and long-range dipolar interactions, have been proposed as an ideal platform for many applications ranging from quantum simulation and information processing to precision measurement~\cite{Micheli2006spintoolbox,Carr2009review,Blackmore2018QuantumReview,augenbraun2020lasercoolingPrecMeas}. These possibilities have led to intense experimental efforts to produce, cool and control molecules, with many advances along the way. In particular, the approach of direct laser-cooling has seen tremendous progress in the past few years. Starting with the first molecular magneto-optical traps (MOTs)~\cite{Barry2014MOT,Truppe2017Mot, Anderegg2017MOT, Collopy2018MOT}, direct laser-cooling promises to be a versatile and efficient route into the ultracold regime for a large variety of molecules~\cite{kozyryev2017sisyphus,Lim2018YbFEDM,mitra2020CaOCH3,Augenbraun2020asymmcooling,baum2021CaOH,alauze2021ultracold}. Many potential applications for molecules require conservative trapping in the absence of resonant light, motivating recent work on magnetic trapping ~\cite{Williams2018MagTrap} and optical dipole trapping using off-resonant light~\cite{Anderegg2018ODT, Langin2021SrFODT,Wu2021YOODT}. To date, attractive optical traps have been used and sub-Doppler cooling has been shown to remain somewhat effective in these traps. This has enabled preparation of samples with record phase-space densities and also high-fidelity detection of optically trapped molecules~\cite{Cheuk2018Lambda, Langin2021SrFODT, Wu2021YOODT}. Nevertheless, the large AC Stark shifts that give rise to trapping can themselves be problematic. For example, differential AC Stark shifts between internal states are thought to limit in-trap laser-cooling temperatures~\cite{Cheuk2018Lambda,caldwell2020sideband,Langin2021SrFODT}, coherence times between rotational states~\cite{burchesky2021rotcoh}, and the effectiveness of shielding molecular samples from inelastic collisions~\cite{Anderegg2021shield}. In attractive optical traps, molecules preferentially occupy regions with high trap light intensity and hence experience near-maximal AC Stark shifts. In contrast, molecules in repulsive traps experience minimal AC Stark shifts, since they reside in areas of minimal light intensity. 

\begin{figure}
{\includegraphics[width=\columnwidth]{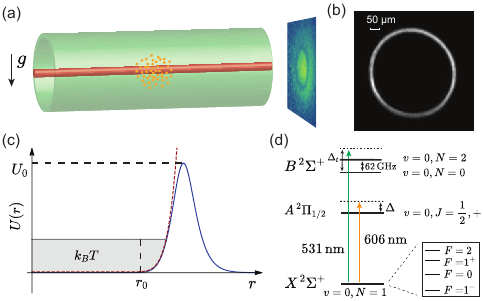}}
\caption{\label{fig1} (a) Experimental Setup. The ring-shaped repulsive optical trap is shown in green. An attractive optical dipole trap (red) formed from a focused Gaussian beam runs along the ring trap axis. Molecules are detected by imaging along the same axis. (b) Intensity distribution of the ring trap with radius set to $r_0 = 160(10)\,\mu\text{m}$. (c) Model potential used to describe the repulsive trap. (d) Energy level diagram of CaF with relevant levels and transitions shown.}
\vspace{-0.2in}
\end{figure}
In addition to reducing differential AC Stark shifts, repulsive optical potentials also aid in preparing large optically trapped samples, crucial for many applications. In experiments to date, the mismatch in trap volume between attractive optical traps (typically $\sim 100\,\mu$m due to limitations in laser power) and the initial MOTs (mm-sized) have limited transfer fractions to well below unity~\cite{Cheuk2018Lambda, Langin2021SrFODT, Wu2021YOODT}. Since repulsive optical potentials can be more power-efficient, larger trap volumes are possible given the same absolute AC polarizabilities. In particular, for an attractive trap formed from a single laser beam, the power required for a given trap depth scales with the cross-sectional area of the laser beam, while for a ring-shaped repulsive trap, the required power scales with the circumference. 

These two benefits, smaller AC Stark shifts and larger trap volumes, motivate exploring trapping of laser-cooled molecules with repulsive optical potentials. In this work, we demonstrate 2D-trapping and laser-cooling of CaF molecules in a near-detuned repulsive optical barrier. The starting point for this work is a $\Lambda$-cooled cloud of CaF molecules in the $\left |X, v=0, N=1\ket$ rotational manifold at zero magnetic field~\cite{Supplement}. In brief, a DC-MOT~\cite{Tarbutt2015DCMOT} of CaF molecules is loaded from a cryogenic buffer gas beam (CBGB)~\cite{Hutzler2012CBGB} that is slowed via chirped slowing~\cite{Truppe2017chirp,Holland2021GSA}. The MOT is subsequently compressed by ramping down the MOT beam powers and ramping up the magnetic field gradient. The magnetic field is then switched off and $\Lambda$-enhanced gray molasses cooling is turned on for $10\,\text{ms}$. The $\Lambda$-cooling light has a single photon detuning of  $\Delta=26\,\text{MHz}$ and contains two hyperfine components addressing the $\left|J=1/2,F=1\ket$ and $\left|J=3/2, F=2\ket$ states. The $\Lambda$-cooled cloud consists of $4.5\times10^4$ molecules at a temperature of $\sim 10\,\mu \text{K}$, and has a Gaussian diameter ($2\sigma$) of 1.2(1)~mm. 

The ring-shaped repulsive optical trapping potential is generated using a laser beam with light blue-detuned by $\Delta_t=108\,\text{GHz}$ from the $X\, ^2\Sigma^+(v=0,N=1) \rightarrow B\,^2 \Sigma^+ (v=0,N=0)$ transition. The ring size is dynamically tunable using two liquid lenses over ms timescales~\cite{Supplement}. To a good approximation, the potential is axially invariant in the region explored by the molecules. In addition, the lack of thermalization over the experimental timescales implies that only the radial dynamics are relevant and the situation is effectively 2-dimensional. Since the molecular temperature $k_B T$ is much lower than the barrier height $U_0$, the potential explored by the molecules is well-approximated by
\begin{equation}
      U(\mathbf{r})=
        \left\{ \begin{array}{ll}
            0, &r\leq r_0 \\
           a (r-r_0)^{\alpha}, &r>r_0,
        \end{array} \right.
    \end{equation}
where $\alpha\gg 1$  (Fig.~\ref{fig1}(c)).

\begin{figure}[t]
	{\includegraphics[width=\columnwidth]{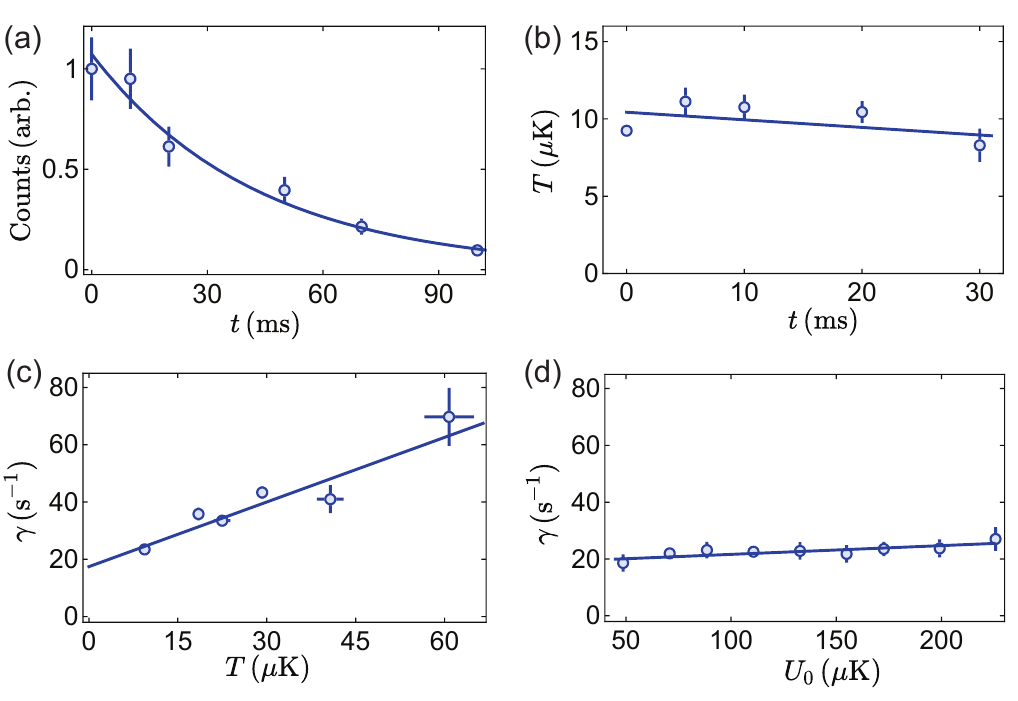}}
	\caption{\label{fig2} (a) Molecular number $N$ versus hold time $t$. The solid line shows an exponential fit, which gives a $1/e$ decay time of $42(3)$\,ms. (b) Temperature $T$ versus hold time $t$. (c) Loss rate $\gamma$ versus initial molecular temperature $T$. (d) Loss rate versus barrier height $U_0$. For (b,c,d), solid lines show linear fits to the data. }
	\vspace{-0.2in}
\end{figure}
In optical traps, undesirable effects such as differential AC Stark shifts and heating are proportional to the trap light intensity. Since the off-resonant photon scattering rate is also proportional to the light intensity, one can use the trap-averaged photon scattering rate $\langle\Gamma_{sc}\rangle_{\text{trap}}$ as a figure-of-merit. For a thermal ensemble in an attractive trap, $\langle\Gamma_{sc}\rangle_{\text{trap}} \propto V$, where $V$ is the trap depth. In contrast, for a repulsive trap, the scaling of $\langle\Gamma_{sc}\rangle_{\text{trap}}$ with barrier height is geometry-dependent. For the ring-shaped barrier explored here, $\langle\Gamma_{sc}\rangle_{\text{trap}}$ is approximately linear with $T$ and independent of barrier height:
\begin{equation}\label{eq:sc}
	\langle\Gamma_{sc}\rangle_{\text{trap}} \propto \frac{\int {d\mathbf{r}} \, U(\mathbf{r})e^{-U(\mathbf{r})/(k_BT)} }{\int {d\mathbf{r}} \, e^{-U(\mathbf{r})/(k_BT)} }\propto T\left(\frac{T}{a}\right)^{\mathcal{O}(1/\alpha)} 
\end{equation}
Photon scattering and therefore differential Stark shifts can therefore be strongly suppressed at low temperatures.
We note in passing that similar scaling laws for other repulsive trap geometries have previously been derived in work with ultracold atoms~\cite{friedman2002dark}. 

To observe these scalings, we first characterize trap heating and loss, which serve as probes for $\langle\Gamma_{sc}\rangle_{\text{trap}}$. Molecules are transferred from the $\Lambda$-cooled cloud into a conventional attractive optical dipole trap (ODT) concentric with the ring trap. The ODT is generated using a single laser beam at 1064\,nm focused to a Gaussian waist of $60(7)\,\mu \text{m}$, much smaller than the repulsive ring radius, and retro-reflected to form a 1D lattice~(Fig.~\ref{fig1}(a)). Subsequently, the cooling light is switched off and untrapped molecules fall away over $50\,\text{ms}$. The molecules are then released into the repulsive ring trap, which has radius $r_0 = 160(10)\,\mu\text{m}$ and a barrier height of $U_0 /k_B = 240(30)\,\mu\text{K}$. Subsequently, a $2\,\text{ms}$ $\Lambda$-cooling pulse recools the cloud to $10(1)\,\mu\text{K}$. After holding the sample in the trap for time $t$, the number and the temperature $T$ are measured. We measure a $1/e$ lifetime of $42(3)$\,ms, but find no observable temperature increase over this timescale (Fig.~\ref{fig2}(a,b)). 

At first sight, the lack of significant heating seems to contradict Eq.~(\ref{eq:sc}), which predicts $\dot{T} \propto T$ and hence exponentially increasing temperatures. Our observations are nevertheless consistent with theory when one takes into account rotational loss due to Raman scattering, which is also proportional to $\langle\Gamma_{sc}\rangle_{\text{trap}}$. Specifically in our case, molecules off-resonantly excited to $\left|B, v=0, N=0\ket$ always return to $\left|X, v=0, N=1\ket$ and experience recoil heating, but molecules excited to $\left|B, v=0, N=2\ket$ are lost from detection if they decay to $\left|X, v=0, N=3\ket$. One therefore expects a rotational loss rate $\sim \bra \Gamma_{sc}\ket_{\text{trap}}$. On the other hand, photon scattering imparts kinetic energy at a rate of $\sim \bra \Gamma_{sc} \ket_{\text{trap}} E_R$, where $ E_R= \hbar^2 k^2/(2m) = k_B \times  0.58\,\mu \text{K}$ is the recoil energy, and $k$ is the trapping light wavevector. Since the initial temperature ($10\,\mu\text{K}$) is well above $E_R/k_B$, molecules are rotationally lost before significant heating occurs. This generically holds true in far-detuned optical traps when temperatures are well above $E_R/k_B$, a regime reached in many laser-cooling experiments.

The observed loss rate is therefore a proxy for $\langle\Gamma_{sc}\rangle_{\text{trap}}$, and should therefore be linear in temperature $T$ and independent of barrier height $U_0$, in accordance with Eq.~(\ref{eq:sc}). To observe the temperature depedence, we vary the molecular temperature between $9.4(3)\,\mu\text{K}$ and $61(4)\,\mu\text{K}$ by adjusting the cooling parameters during a 2\,ms $\Lambda$-cooling pulse following trap loading. Indeed, we find that the $1/e$ loss rate $\gamma$ increases linearly with $T$ with an offset at $T=0$ (Fig.~\ref{fig2}(c)). The $T=0$ offset could arise from residual light on the interior of the ring and other loss mechanisms such as collisions with background gas and leaked resonant light. We note that the observed loss rates are $\sim 10^{-2}\,\Gamma_{sc,\text{max}}$, where $\Gamma_{sc,\text{max}}$ is the theoretically predicted loss rate at the peak barrier intensity~\cite{Supplement}. Since the average AC Stark shifts are also proportional to $\bra \Gamma_{sc}\ket_{\text{trap}}$, this indicates that they are strongly suppressed compared to an attractive trap with a similar depth and absolute polarizability.

We next probe the dependence of $\bra \Gamma_{sc}\ket_{\text{trap}}$ on barrier height $U_0$. We begin with molecules at $10(1)\,\mathrm{\mu K}$ within the barrier, and then suddenly change the barrier height $U_0/k_B$ to a value between $50\,\mathrm{\mu K}$ and $230\,\mathrm{\mu K}$. As shown in Fig.~\ref{fig2}(d), the loss rates increase slightly with $U_0$. This generally agrees with Eq.~(\ref{eq:sc}), which predicts $\gamma$ to be largely independent of $U_0$, in stark contrast to the linear scaling found in attractive traps ($\gamma \propto V$). The small increase in loss rate with $U_0$ likely arises from residual light within the ring. 

Having explored the dependences of loss and heating on $T$ and $U_0$, we next investigate whether laser-cooling, specifically $\Lambda$-cooling, continues to be effective within the repulsive barrier. Because of the suppressed AC Stark shifts, one could expect in-trap laser-cooling to perform similarly as in free space, in contrast to attractive optical traps where significant AC Stark shifts can affect the effectiveness of $\Lambda$-cooling by destabilizing the coherent dark states involved. This should be true on timescales shorter than the transit time $\tau_t$ of a molecule across the trap ($\tau_t  = r_0/\sqrt{2k_BT/m} \approx 1-3\,\text{ms}$). On longer timescales, the effectiveness of in-trap $\Lambda$-cooling is not guaranteed. The repulsive trap light can lead to heating and loss in several ways. First, the electronically excited state could experience opposite AC Stark shifts and the molecules could become untrapped when they reach the ring. Second, strong differential Stark shifts could convert $\Lambda$-cooling into $\Lambda$-heating. In our case, these are non-negligible, since $\Delta_t \approx 5 B$, where $B\approx 20\,\text{GHz}$ is the rotational constant. Third, multi-photon processes involving the repulsive trap light in combination with the cooling light can also lead to heating or loss. 

\begin{figure}
	{\includegraphics[width=\columnwidth]{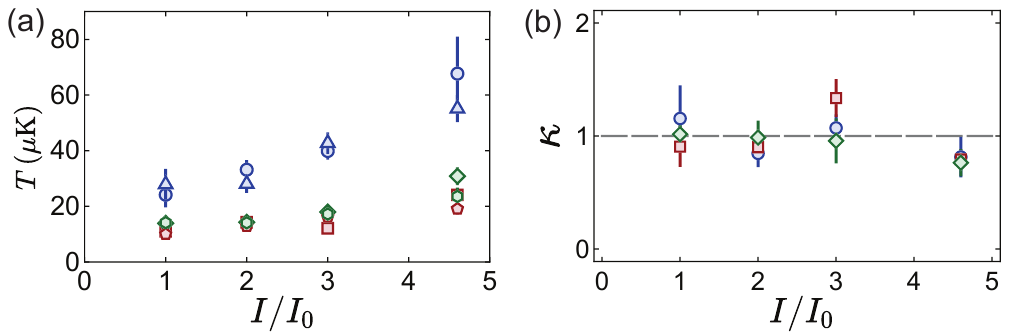}}
	\caption{\label{fig3} (a) Temperature versus cooling light intensity $I/I_0$ at various $\Lambda$-cooling light detunings $\Delta$, in free space and in-trap. Blue circles, green diamonds, and red squares show free space temperatures; Blue triangles, green pentagons, and red hexagons show in-trap temperatures. (b) Ratio of the in-trap to free-space temperature $\kappa$ as a function of cooling light intensity $I/I_0$ at various detunings. For both plots: $I$ is the single-beam single-axis intensity,  $I_0=5.0(5)\,\text{mW/cm}^2$; $\Delta = 6$\,MHz (blue circles and  triangles), 13\,MHz (green diamonds and pentagons), and 38\,MHz (red squares and hexagons). }
\vspace{-0.2in}
\end{figure}

To characterize in-trap $\Lambda$-cooling, we compare temperatures obtained in free space to those obtained in-trap following identical laser-cooling pulses. Molecules are loaded from the $\Lambda$-cooled cloud into the ring trap by suddenly switching the trap on, which ensures that the in-ring molecules are in contact with the repulsive light. Subsequently, we wait $30\,\text{ms}$ for untrapped molecules to fall away. A $5\,\text{ms}$ $\Lambda$-cooling pulse with various detunings $\Delta$ and intensities $I$ is then applied, both with or without the ring present. The temperature is then measured via time-of-flight expansion. Over the explored range of $\Lambda$-cooling parameters, we observe no significant difference between temperatures obtained in-trap ($T_{R}$) and those obtained in free space ($T_{FS}$) (Fig.~\ref{fig3}(a)). Notably, the minimum temperature reached in the trap is similar to that obtained in free-space. To investigate whether additional heating processes involving trap light are present, we plot the ratio $\kappa= T_{R}/T_{FS}$ as a function of the cooling light intensity $I$ (Fig.~\ref{fig3}(b)), and observe no dependence. This rules out any significant heating contributions from processes involving both cooling light and trapping light, since an incoherent process with $n$ photons of cooling light would show $I^n$ dependence.

We next examine whether laser-cooling leads to additional losses for the trapped molecules. We compare the $1/e$ lifetimes $\tau$ of the trapped molecules with and without $\Lambda$-cooling light, and find $\tau=\,57(1)\,\text{ms}$ and $\tau=40(1)\,\text{ms}$ respectively. These are much longer than the transit time $\tau_t$, which indicate that in-trap cooling does not lead to additional losses compared to trapping alone. We note in passing that we obtain lifetimes up to $100\,\text{ms}$ when the repulsive barrier light is further detuned~\cite{Supplement}.

\begin{figure}
	\includegraphics[width=\columnwidth]{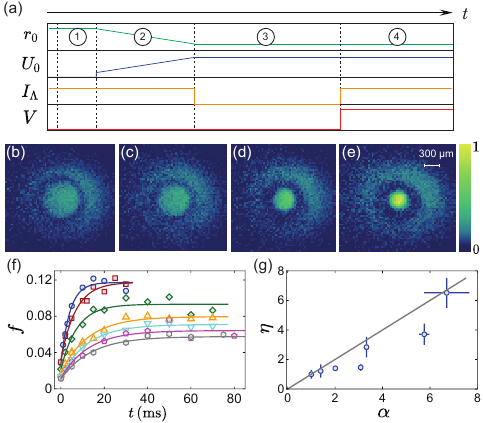}	
	\caption{\label{fig4} (a) Experimental sequence for enhanced transfer into an attractive ODT. This consists of four steps: (1) free-space cooling (10\,ms), (2) compression (21\,ms), (3) release of untrapped molecules (30\,ms), and (4) transfer into the attractive ODT. Shown are the ring radius $r_0$, barrier height $U_0$, cooling light intensity $I$, and attractive ODT depth $V$ as a function of time. (b,c,d,e) In-situ images during compression, taken 0, 6, 15, 21\,ms, into the ramp, respectively. (f) Transfer fraction $f$ versus loading time $t$. From top to bottom, the final trap radii are $r_f= 160(10), 172(3), 227(2), 292(2), 350(1), 414(2)\,\mu\text{m}$, respectively. Solid lines show fits to an exponential saturation curve. (f) Loading rate enhancement $\eta$ versus compression ratio $\alpha$. Solid line indicates $\eta = \alpha$, the expected geometric enhancement.  }
\vspace{-0.2in}
\end{figure}

In the last part of this work, we illustrate how repulsive optical traps can be used to transfer molecules from a MOT into an optical trap with high efficiency. In past experiments, the size mismatch of typical optical traps with MOTs have been a limiting factor in achieving high transfer fractions. Although $\Lambda$-cooling-assisted optical loading helps by enlarging the capture volume beyond the bare size of the optical trap, the highest observed transfer fractions have been limited to $\sim 5\%$ to date~\cite{Langin2021SrFODT}. As mentioned earlier, repulsive traps with a suitable geometry are more power-efficient than their attractive counterparts, and can therefore be made much larger. To illustrate this, we set the repulsive ring radius to $r_0= 414(2)\,\mathrm{\mu m}$ (barrier height $U_i/k_B = 65(1) \,\mu\text{K}$) to maximize the spatial overlap with the $\Lambda$-cooled cloud, which is similar in size to the initial MOT. The ring trap is switched on following initial $\Lambda$-cooling, after which we wait 21\,ms for any transient dynamics to damp out. We subsequently image the molecular cloud in-situ using a $250\,\mu$s imaging pulse with resonant light. Despite blurring from the imaging process~\cite{Supplement}, we observe a clear boundary between trapped and untrapped molecules, and find that $27(3)\%$ of the molecules are captured into the repulsive trap~(Fig.~\ref{fig4}(b)).  

Despite its advantage of achieving large trap volumes with limited laser power, the ring-shaped geometry has a drawback. Since the molecule density is set by the ring size, in-trap laser-cooling produces negligible density enhancement, unlike in Gaussian-shaped attractive traps where density is strongly enhanced at lower temperatures~\cite{Anderegg2018ODT, Cheuk2018Lambda, Langin2021SrFODT,Wu2021YOODT}. To benefit from both the large capture volume of the repulsive trap and the density enhancement offered by a conventional attractive ODT, we implement the following two-step transfer scheme. After initial transfer into the large volume ring trap, we compress the trap in the presence of laser-cooling by dynamically tuning the ring size. The compressed samples are then loaded into an attractive ODT. With better mode-matching offered by trap compression, high transfer fractions can be achieved.

In detail, we switch on the repulsive ring with initial radius $r_i = 414(2)\,\mu\text{m}$ and barrier height $U_i/k_B = 65(1) \,\mu\text{K}$. Subsequently, the radius is smoothly compressed to $r_f = 160(10) \,\mu \text{m}$ and the barrier height is ramped to $U_f/k_B=  240(30)\,\mu\text{K}$ over $21\,\text{ms}$ (Fig.~\ref{fig4}(a))~\cite{Supplement}. In-situ images show rising densities throughout compression (Fig.~\ref{fig4}(b,c,d,e)). In addition, we find that $T$ remains constant at its initial value of $T_i\approx 10\,\mu\text{K}$~\cite{Supplement}. This indicates sufficient cooling, since in the absence of cooling, compression-induced heating would lead to a final temperature  $T_f\geq \alpha T_i$, where the lower bound is for an adiabatic ramp and $\alpha = r_i^2/r(t)^2$ is the compression ratio.

To transfer molecules from the repulsive trap into the much smaller attractive ODT, we switch on the attractive trap in the presence of $\Lambda$-cooling light. Fig.~\ref{fig4}(f) shows the transfer fraction $f=N_{\text{ODT}}/N_{\text{SD}}$ as a function of transfer time $t$, where $N_{\text{SD}}$ is the initial molecule number in the $\Lambda$-cooled cloud measured within the same experimental sequence. The transfer fraction initially rises and then saturates, with both the loading rate and the saturated transfer fraction increasing with compression. Since the initial loading rate $R_0=\dot N_{\text{ODT}}(t=0)/N_{\text{SD}}$ is proportional to the initial density, the normalized initial loading rate $\eta = R_0(\alpha)/R_0(\alpha=1)$ directly measures the density enhancement. As shown in Fig.~\ref{fig4}(g), we observe a peak enhancement factor of $6$ and find that $\eta \approx \alpha$, consistent with the geometric expectation from the reduced ring area. At maximum compression ($\alpha\approx 6$), rapid saturation times around $10 \,\text{ms}$ are observed, much faster than the $\sim 100\,\text{ms}$ times previously reported~\cite{Cheuk2018Lambda,Wu2021YOODT,Langin2021SrFODT}. We also observe highly efficient transfer, with $45(5)\%$ of the repulsively trapped molecules transferred into the attractive ODT. This corresponds to an overall transfer efficiency of 12(2)\% from the $\Lambda$-cooled cloud, and a total of $5.4(5)\times10^{3}$ molecules in the attractive ODT.

Although the loading rate  is enhanced six-fold by trap compression, the final trapped number is enhanced two-fold. We believe that the overall transfer efficiency is likely limited by the number initially captured in the repulsive trap. We estimate that the transfer fraction is consistent with the observed lifetimes in the repulsive trap, which can be increased with faster compression. Theoretically, we find that the compression rate $\dot{r}/r$ can in fact be as fast as the cooling rate $\dot{T}/T$~\cite{Supplement}. Based on the observed and theoretically predicted sub-Doppler cooling timescales of $\approx 100\,\mu\text{s}$~\cite{devlin2016coolingSim,Anderegg2018ODT,Caldwell2019singlefreq,Ding2020YOSD}, sub-ms-scale compression, much faster than the 21\,ms used here, is in principle possible. 

In conclusion, we have explored trapping and laser-cooling of molecules in a near-detuned repulsive optical potential. We find rotational loss, rather than heating, to be the dominant feature at laser-cooling temperatures. Loss measurements reveal temperature and power dependences characteristic of repulsive traps, consistent with strongly suppressed off-resonant photon scattering rates and average AC Stark shifts. In addition, despite using repulsive light detuned by only a few rotational spacings, we find that in-trap $\Lambda$-cooling performs similarly as in free space, and find no evidence of additional losses due to the trapping light. By combining laser-cooling with dynamical reshaping of a repulsive trap, we have also presented a new method that rapidly transfers laser-cooled molecules into optical traps with record efficiencies. With improvements such as faster compression and more optimal trap shaping, our method could allow nearly lossless transfer of molecules from MOTs into optical traps, overcoming an outstanding challenge in experiments to date. 

Looking ahead, laser-cooling in dynamically tunable repulsive traps could be useful in a variety of future explorations, such as evaporative cooling. The shielding of molecules from collisional loss is expected to be much more effective due to suppressed differential Stark shifts~\cite{Anderegg2021shield}, and dynamical compression can offer high starting densities. The work here could also guide the development of bottle-beam optical tweezers~\cite{Barredo2020bottle} for molecules, where internal state decoherence caused by differential AC Stark shifts could be suppressed~\cite{burchesky2021rotcoh}. Methods shown in this work could also be useful for precision measurement experiments with trapped and laser-cooled molecules~\cite{augenbraun2020lasercoolingPrecMeas}. Repulsive ring traps could suppress effects due to trap inhomogeneity, and dynamic trap decompression could be used to lower molecular densities and temperatures, thereby reducing systematic effects that arise from Doppler shifts and molecular interactions.
\vspace{0.2in}

Y.L. and C.M.H. contributed equally to this work.

\bibliographystyle{apsrev4-1}
%


\end{document}


\title{Supplemental Material for ``Molecular Laser-Cooling in a Dynamically-Tuned Repulsive Optical Trap" }
\affiliation{Department of Physics, Princeton University, Princeton, NJ 08544 USA}
\affiliation{Department of Electrical and Computer Engineering, Princeton University, Princeton, NJ 08544 USA}
\author{Yukai Lu}
\affiliation{Department of Physics, Princeton University, Princeton, NJ 08544 USA}
\affiliation{Department of Electrical and Computer Engineering, Princeton University, Princeton, NJ 08544 USA}
\author{Connor M. Holland}
\affiliation{Department of Physics, Princeton University, Princeton, NJ 08544 USA}
\author{Lawrence W. Cheuk}
\affiliation{Department of Physics, Princeton University, Princeton, NJ 08544 USA}
\date{\today}
\maketitle

\section{Experimental Apparatus and Preparation of Molecular Samples}
\subsection{Apparatus Overview}
\begin{figure}
	{\includegraphics[width=\columnwidth]{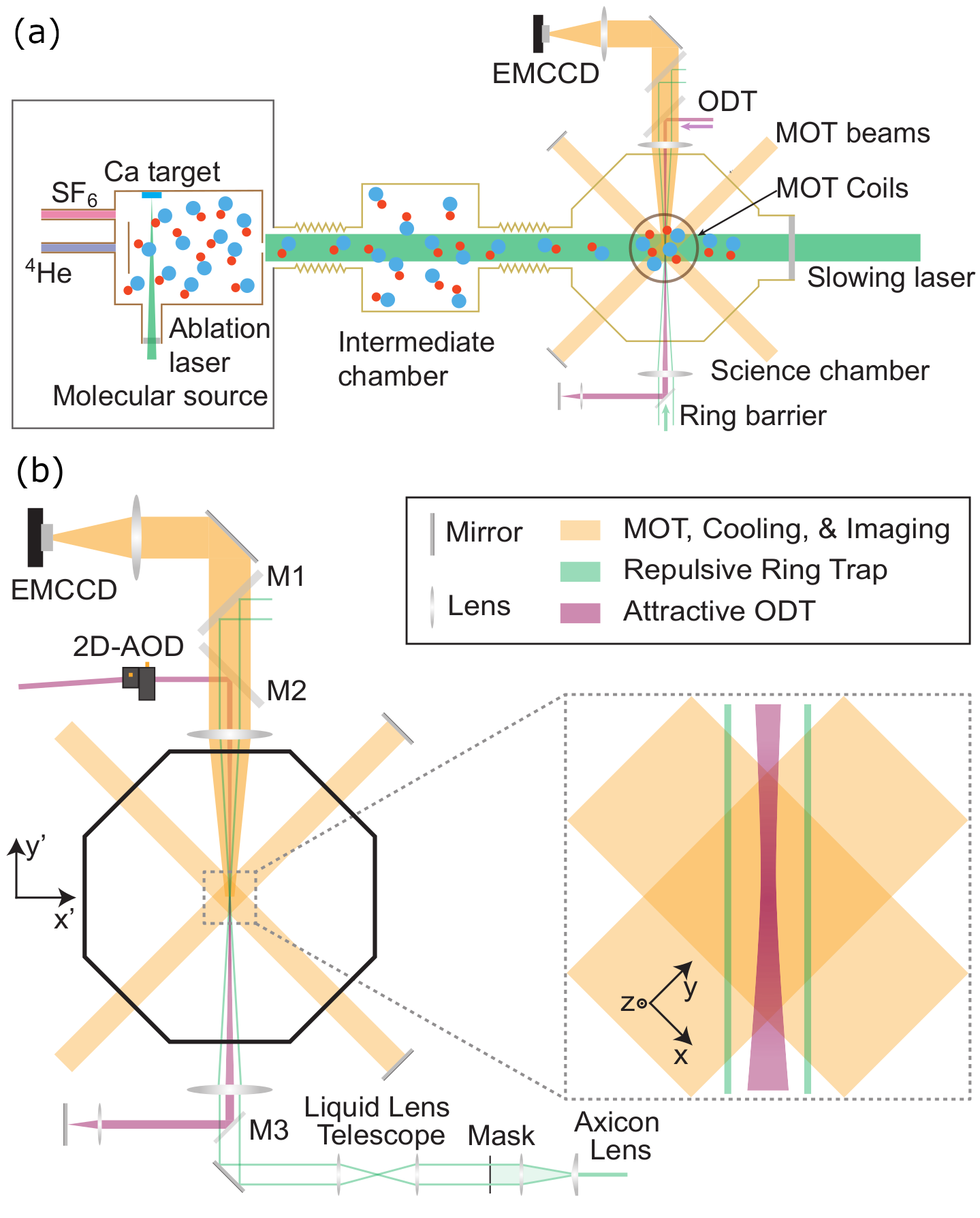}}
	\caption{\label{fig:apparatus} (a) Apparatus Schematic. (b) Experimental layout. Shown in orange are laser beams for magneto-optical trapping, cooling, and imaging. The repulsive ring trap and the attractive ODT are shown in green and magenta, respectively. They are concentric and pass through the center of the chamber where the molecules are positioned. M1, M2 and M3 are are dichroic mirrors used to combine and split the relevant beams. For detection, we collect fluorescence onto an EMCCD camera along the imaging axis ($y'$). }
\end{figure}

Our apparatus consist of three parts, a cryogenic chamber used to produce a molecular beam, an intermediate chamber, and a main chamber where laser-cooling and trapping occurs (Fig.~\ref{fig:apparatus}(a)). The three chambers are connected in series by differential pumping tubes. During the experiment, the cryogenic chamber has a typical pressure of $\sim 1\times10^{-9}$\,Torr, while the intermediate and main chambers are held at ultra high vacuum pressures (UHV) of $\sim3\times10^{-10}$\,Torr and $\sim1\times10^{-10}$\,Torr, respectively.

Laser beams used for laser-slowing, magneto-optical trapping, sub-Doppler cooling and fluorescent imaging enter the chamber from a variety of paths. The main cooling light ($X\, ^2\Sigma^+(v=0,N=1) \rightarrow A\, ^2\Pi_{1/2}(v=0, N=0)$) used for magneto-optical trapping and $\Lambda$-cooling/imaging enter along the $x$,\,$y$,\,$z$ axes. Also, the first vibrational repumping light ($X\, ^2\Sigma^+(v=1,N=1) \rightarrow A\, ^2\Pi_{1/2}(v=0, N=0)$) is also sent along these paths. The slowing light ($X\, ^2\Sigma^+(v=0,N=1) \rightarrow B\, ^2\Sigma^+(v=0, J=1/2,+)$) and three vibrational repumpers ($X\, ^2\Sigma^+(v=1,2,3,N=1) \rightarrow A\, ^2\Pi_{1/2}(v=0, N=0)$) enter along the molecular beam axis $x'$, which is oriented 45 degrees to the $x$,\,$y$ axes (Fig.~\ref{fig:apparatus}(b)). 

For our work here, we detect molecules by measuring their fluorescence as they cycle on the $X$-$A$ transition. This fluorescence is imaged along the imaging path ($y'$ axis) onto an EMCCD camera. The repulsive ring trap propagates along the imaging path and is mostly filtered out with a dichoric mirror before reaching the camera. The attractive optical dipole trap is also sent in along this same axis using a dichroic mirror, but propagates opposite to the imaging path.

\subsection{Molecular Source}
Our experimental sequence begins with the generation of a cryogenic buffer gas beam (CBGB) of CaF molecules with significant population in the laser-coolable $X\, ^2\Sigma^+ (v=0, N=1)$ manifold~\cite{Hutzler2012CBGB}. In detail, a Ca target is ablated using a pulsed Nd:YAG laser (frequency doubled to $532\,\text{nm}$) in the presence of $^4$He and SF$_6$ inside a buffer gas cell held at $3.5\,\text{K}$. A subsequent chemical reaction produces CaF molecules~\cite{Truppe2018CaFbeam}, which are cooled via collisions with the $^4$He gas, producing a bright beam of CaF molecules with a mean forward velocity of 150 to $180\,\text{m/s}$. We typically flow $1\,\text{sccm}$ of $^4$He and $0.05\,\text{sccm}$ of SF$_6$ into the cell. 

\subsection{Chirped Laser Slowing of Molecular Beam}
The CBGB of molecules are then slowed via chirped slowing to near zero velocity using a counter-propagating laser beam addressing the rotationally closed $X\, ^2\Sigma^+ (v=0, N=1) \rightarrow B\,^2\Sigma^+ (v=0,N=0)$ transition ($\lambda=531.0\,\text{nm}$)~\cite{Truppe2017chirp}. To address the four hyperfine manifolds in order to obtain photon cycling, we engineer the required frequency spectrum using a high-bandwidth fiberized electro-optical modulator (EOM) combined with a computer generated holography (CGH) technique~\cite{holland2021gsa}. To slow molecules, the frequency of the cooling light is continuously chirped to stay on resonance with decelerating molecules. The slowing light is turned on 2 to 3\,ms following laser ablation, and is initially red-detuned by $\approx500\,\text{MHz}$. Its frequency is then ramped linearly to resonance over $4.7\,\text{ms}$. This sequence maximizes the slow molecular population (within the MOT capture velocity) at the center of our UHV chamber, which is centered $80\,\text{cm}$ from the exit aperture of our buffer gas cell.

To repump molecules that vibrationally decay into the $v=1,2,3$ manifolds during the slowing process, we combine our slowing light with three vibrational repumpers addressing the $X\, ^2\Sigma^+ (v=1, N=1) \rightarrow A\, ^2\Pi_{1/2} (v=0, J=1/2, +)$ ($\lambda=628.6\,\text{nm}$), $X\, ^2\Sigma^+ (v=2, N=1) \rightarrow A\, ^2\Pi_{1/2} (v=1, J=1/2, p=+)$ ($\lambda  = 628.1\,\text{nm}$), and $X\, ^2\Sigma^+ (v=3, N=1) \rightarrow A\, ^2\Pi_{1/2} (v=2, J=1/2, p=+)$ ($\lambda = 627.7\,\text{nm}$) transitions. The frequencies of the $v=1,2$ repumpers are also ramped during chirped slowing. Frequency sidebands addressing the ground state hyperfine levels are imprinted using free-space resonant EOMs at a modulation frequency of $f=25\,\text{MHz}$ with modulation indices of $\beta=3.2(2)$ and $\beta=3.8(2)$ for $v=1$ and $v=2,3$ repumpers, respectively. The slowing beam and repumping beams all have diameters of $\sim 10\,\text{mm}$, and their powers are 880\,mW for slowing, 290\,mW for $v=1$, 35\,mW for $v=2$, and 3\,mW for $v=3$. 

\subsection{Magneto-Optical Trapping and Compression}
The slowed molecules are then loaded into a DC magneto-optical trap (MOT) operating on the $X\, ^2\Sigma^+ (v=0, N=1) \rightarrow A\, ^2\Pi_{1/2} (v=0, J=1/2, +)$ transition ($\lambda = 606.3 \text{nm}$)~\cite{Tarbutt2015dcMOT}. The MOT light is red-detuned from resonance ($\Delta=-5.5\,\text{MHz}$) and sidebands addressing the four ground state hyperfine components are generated using acousto-optical modulators (AOMs). In addition to the main MOT cooling light, the MOT beams also contain $v=1$ light. The MOT beams are sent into the chamber along 3 orthogonal axes ($x'$, $y'$ and $z$) and retro-reflected. During MOT loading, the $v=1$ light along the slowing path is turned off, but $v=2$ and $v=3$ light remain on.

During slowing, the axial magnetic field gradient is initially set to 26\,G/cm, and the MOT beams are kept on with a power of 22\,mW (single-beam, single-axis). Initially, each MOT beam contain hyperfine components with single-axis single-pass powers of 9.4\,mW, 5.5\,mW, 1.8\,mW, 5.3\,mW for $F=2$, $F=1^+$, $F=0$, and $F=1^-$, respectively, and has a beam diameter of $\sim 1\,\text{cm}$. The $v=1$ repump light has a single-axis single-pass power of 41\,mW. After loading molecules into the MOT for 5\,ms, we compress the MOT by ramping the magnetic gradient to 106\,G/cm over 10\,ms, while simultaneously lowering the MOT beam power from 22\,mW to 2\,mW. 
The power balance between the hyperfine components is unchanged during MOT compression.

\subsection{ $\Lambda$-Enhanced Gray Molasses Cooling}
After the MOT is compressed, the magnetic field is switched off, and we perform $\Lambda$-enhanced gray molasses cooling~\cite{Cheuk2018Lambda}. The MOT frequency is jumped to a single-photon detuning of $\Delta= 21\,\text{MHz}$, the $F=1^+,0$ components are turned off, and the $F=2,1^-$ components are set to be on two-photon resonance. The $v=1,\,2,\,3$ repumpers remain unchanged from the MOT configuration. For optimal $\Lambda$-cooling, we null the magnetic field using three pairs of bias coils.  

To optimize the density and temperature of the molecular samples, we specifically use the following sequence. Molecules are first cooled for 1\,ms at a single-photon detuning $\Delta= 21\,\text{MHz}$, hyperfine ratio $R_{2,1}=P_{F=2}/P_{F=1^-}=0.66$, and total single-axis single-pass power of $P_\Lambda= 9.75 \,\text{mW}$. These values are then ramped over the next ms to the final configuration of $\Delta =26\,\text{MHz}$, $R_{2,1}=0.67$, $P_\Lambda=0.94\,\text{mW}$ and are held constant for the subsequent 8\,ms. This procedure generates a molecular cloud containing $\sim4.5\times10^{4}$ molecules at 10(1)\,$\mu$K with a Gaussian diameter ($2\sigma$) of $1.2(1)\,\text{mm}$. This is the starting point of the work described in the main text.

\section{Generating Attractive and Repulsive Optical Potentials}
\subsection{Attractive Optical Dipole Trap (ODT)}
We form a single-beam attractive optical dipole trap (ODT) using $\lambda=1064\,\text{nm}$ light. The light passes through an 2-axis acousto-optical deflector (2D-AOD), which allows fast switching, location control, and intensity stabilization of the trapping beam. The single trap beam has a maximum power of 25\,W and is focused to a waist of $60(7)\,\mu\text{m}$, providing a single-pass trap depth of $V_0/k_B=170(30)\,\mu\text{K}$. To maximize the total trap depth, the beam is retro-reflected to form a 1D lattice. 

\subsection{Repulsive Ring Trap}
The repulsive ring trap is generated using light blue-detuned from the $X\, ^2\Sigma^+ (v=0, N=1) \rightarrow B\,^2\Sigma^+ (v=0,N=0)$ transition ($\lambda = 531\, \text{nm}$). This light is generated via a DBR seed laser at $\lambda = 1061.8 \,\text{nm}$, which is amplified and frequency doubled. We shape this light into a variable-size ring beam with the setup shown in Fig.~\ref{fig:apparatus}(b). In detail, a ring beam is first generated using an axicon lens, and subsequently passes through a chrome ring mask that is imaged using a telescope with variable magnification onto the molecules. The telescope comprises a pair of liquid lenses (Optotune EL-10-30-C-VIS-LD) whose focal lengths can be adjusted from $f=200\,\text{mm}$ to 100\,mm over ms timescales.

\subsection{Ring Trap Compression Sequence}
\begin{figure}
	\includegraphics[width=\columnwidth]{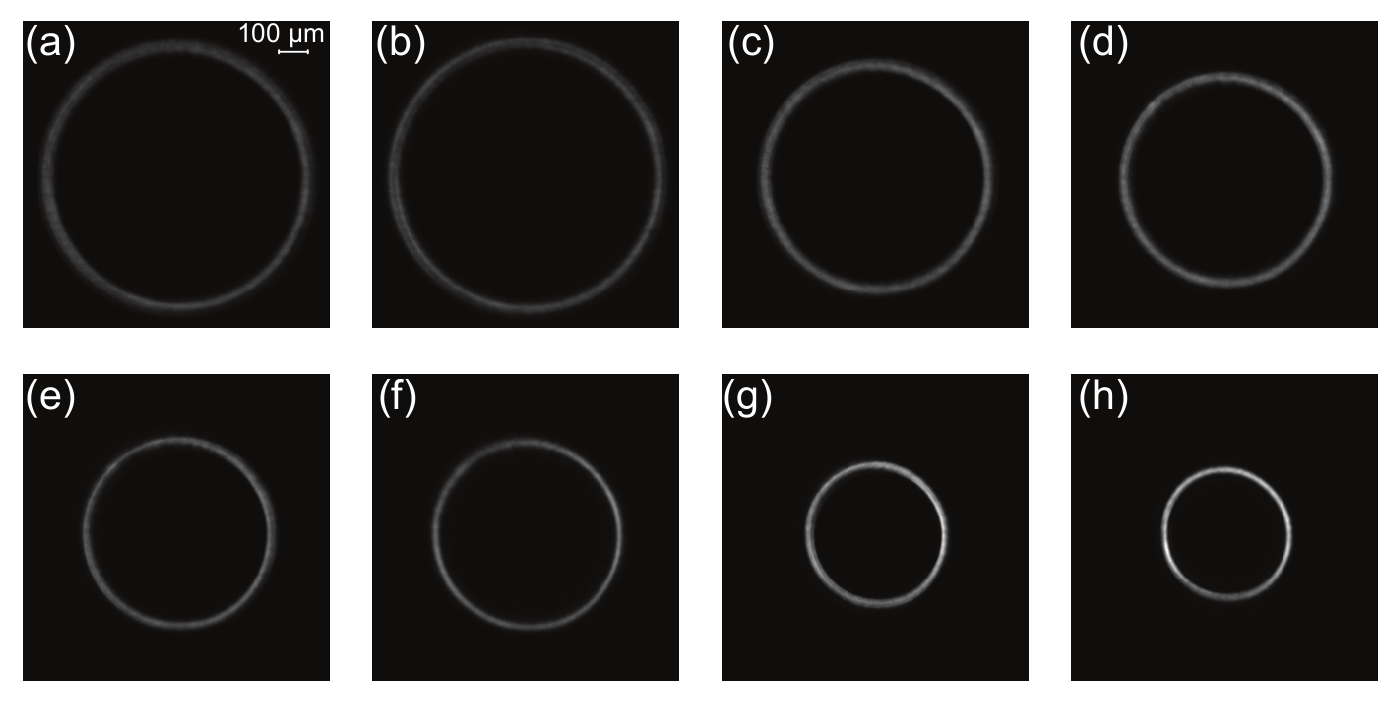}
	\caption{\label{fig:ringpanel} Exemplary images of the ring barrier intensity at various points in the compression sequence. (a-h) are taken 0,\,3,\,6,\,9,\,12,\,15,\,18,\,21\,ms into the compression sequence, respectively. }
\end{figure}

\begin{figure}
\includegraphics[width=0.8\columnwidth]{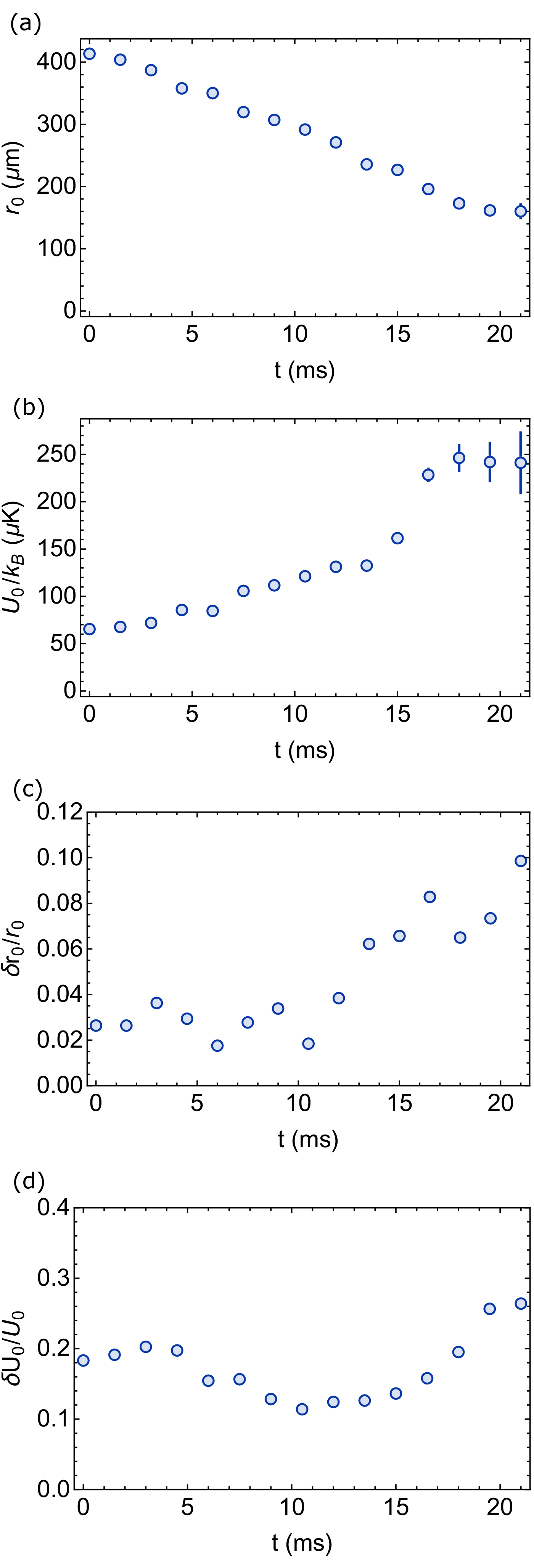}
	\caption{\label{fig:compression} (a) Evolution of mean radius $r_0$ as a function of time $t$ along the compression sequence. (b) Evolution of mean barrier height $U_0/k_B$ as a function of time $t$ along the compression sequence. (c) Fractional variation (standard deviation) of radius $\delta r_0/r_0$ as a function of time $t$ along the compression sequence. (d) Fractional variation (standard deviation) of the barrier height $\delta U_0/U_0$ as a function of time $t$ along the compression sequence.}
\end{figure}

The compression sequence used in this work has a duration of $21\,\text{ms}$. Exemplary images showing the intensity distributions are shown in Fig.~\ref{fig:ringpanel}. The evolution of the ring radius, the barrier height, and their respective fractional variations are shown in Fig.~\ref{fig:compression} as a function of time along the compression sequence. During the compression sequence, the mean ring radius decreases from 413(2)\,$\mu \text{m}$ to 160(10)\,$\mu \text{m}$ while its fractional standard deviation $\delta r_0/r_0$ increases from $2\%$ to $10\%$. The mean barrier height increases from $U_0/k_B = 65(1) \,\mu\text{K}$ to $U_0/k_B = 230(40) \,\mu\text{K}$  while its fractional standard deviation $\delta U_0/U_0$ varies between $10\%$ and $25\%$.  Exemplary images are shown at various times along the compression sequence in Fig.~\ref{fig:ringpanel}.

\subsection{Ring Trap Characterization}\label{sec:ringcharacterization}
To characterize the repulsive ring potential, we directly measure the trap light intensity distribution using a camera placed at an equivalent imaging plane. The camera is triggered at different points during the ring compression sequence, allowing us to also characterize any transient features. 

Since the ring shape slightly deviates from circular due to ring-generation imperfections, we use the following angular averaging procedure to obtain a mean radius $r_0$. From each image, we produce a set of images with various rotation angles uniformly covering $180^\circ$ at a step size of $2^\circ$. For each rotation angle, we extract the average intensity along a vertical strip $14\,\mu\text{m}$ about the ring's center. The 1D intensity distribution is then separated into a left half and a right half, and each is fit to the model potential in the main text. The mean of the two fitted values of $r_0$ provide a mean $r_0$ for this angle. This extracted value is then averaged across all angles to obtain a final mean radius $r_0$. 

\begin{figure}
\includegraphics[width=0.8\columnwidth]{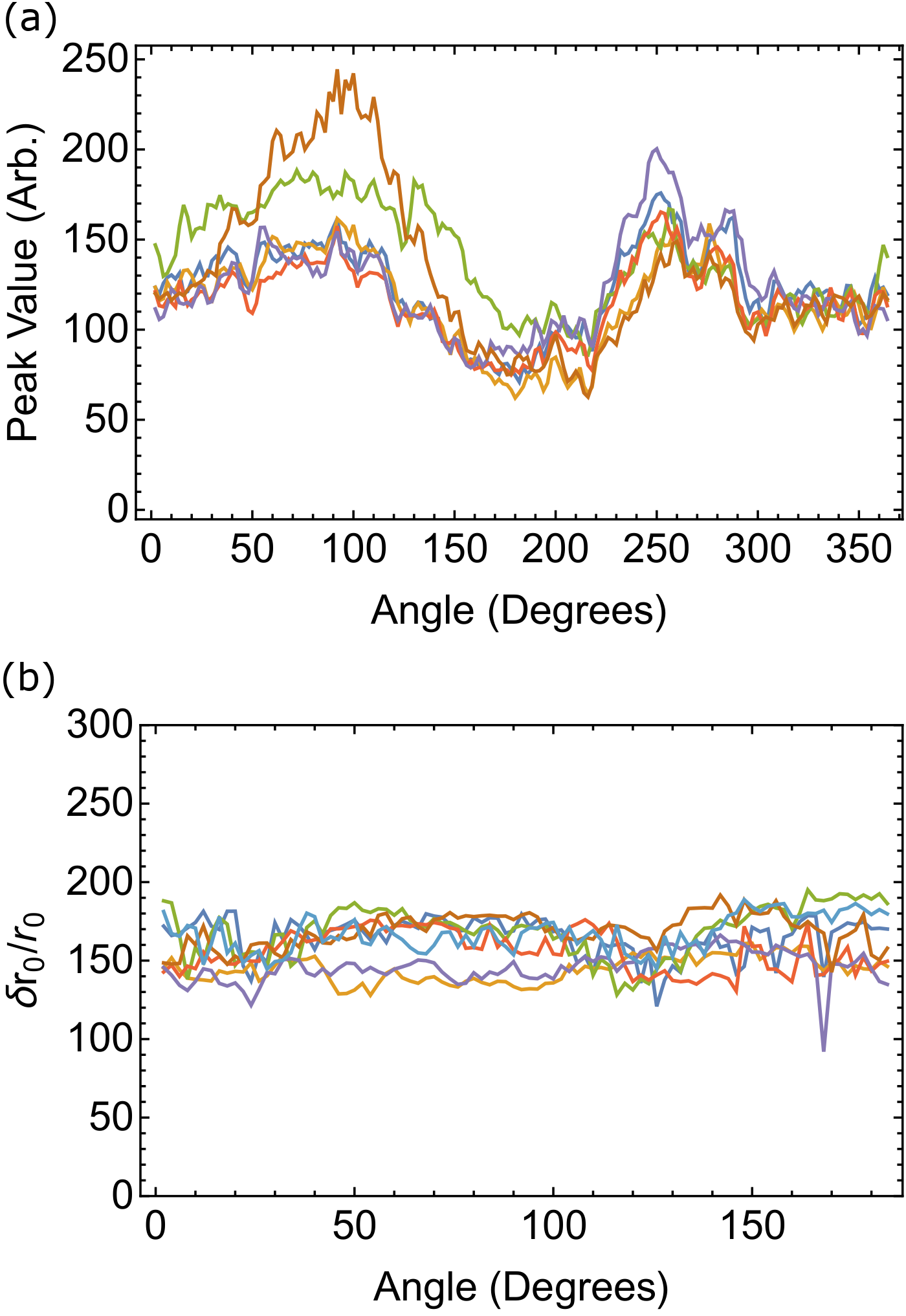}
	\caption{\label{fig:angavg} (a) Exemplary curves showing maximum average counts versus rotation angle for a ring size of 160(10)\,$\mu\text{m}$ for a few measurements. (b) Exemplary curves showing extracted radius versus rotation angle for a ring size of 160(10)\,$\mu\text{m}$.}
\end{figure}

To estimate the barrier height, we discard the left half of the 1D distribution from above, and the maximum intensity is recorded. We then extract the maximum intensity versus rotational angle, which is then converted to a barrier height via the theoretically calculated polarizability. Typically, we observe $10\%$ to $25\%$ variation $\delta U_0/U_0$ (fractional standard deviation) as a function of angle (Fig.~\ref{fig:angavg}). The mean barrier height reported in the main text is the average across all angles, while the error bar is obtained from the variation of this mean value across multiple measurements.

\subsection{Residual Light within Ring Trap}
\begin{figure}
	\includegraphics[width=0.8 \columnwidth]{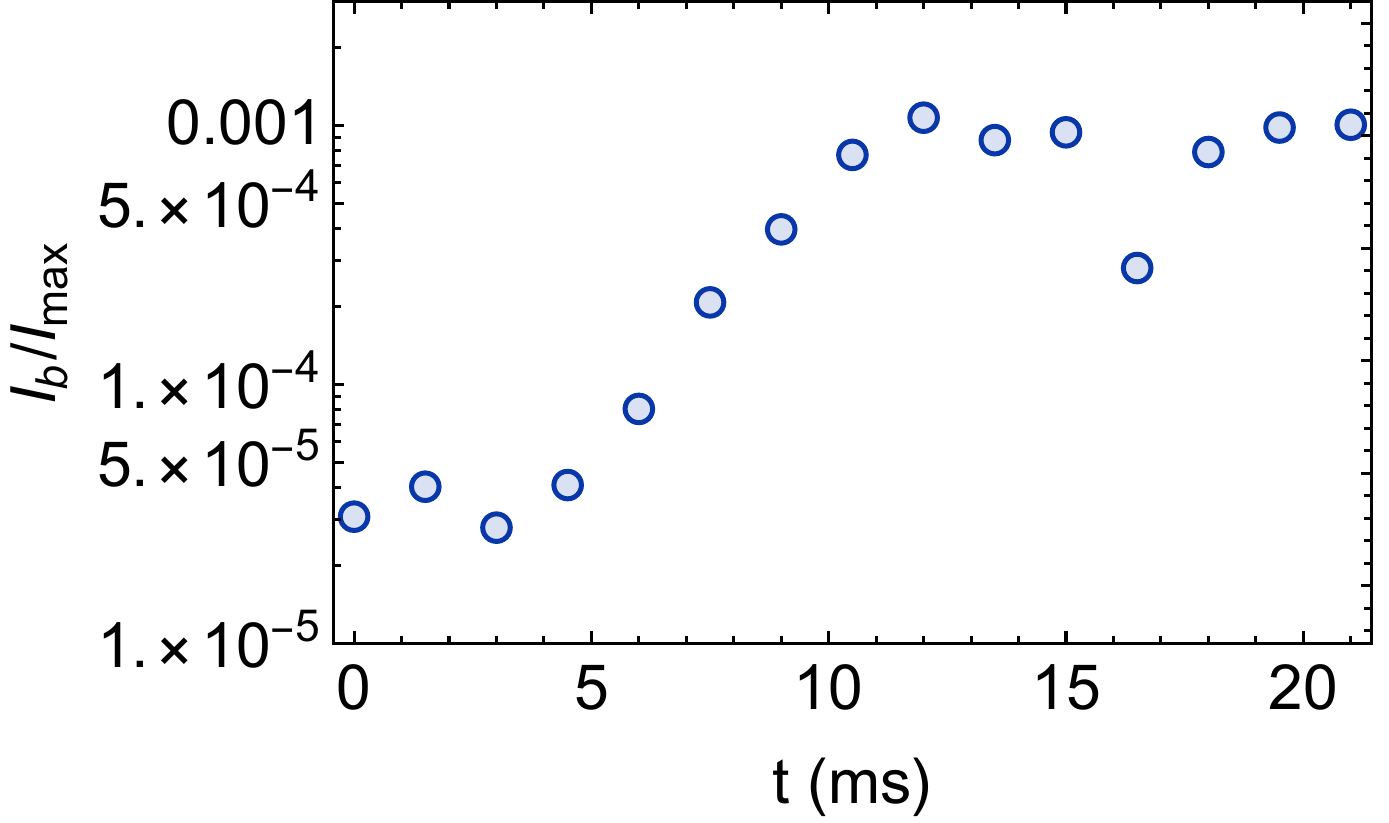}
	\caption{\label{fig:reslight} Suppression ratio $I_b/I_{\text{max}}$ of the average intensity within the ring compared to the peak intensity as a function of time $t$ along the compression sequence.}
\end{figure}

The scaling laws we derive and refer to in the main text assume that the interior of the ring $r\leq r_0$ is perfectly dark. In practice, optical imperfections lead to light in the interior of the ring. To set a lower bound on the leaked light, we measure the average intensity $I_b$ in the interior of the trap $r<0.9\,r_0$, normalized to the average maximum intensity $I_\text{max}$ along the barrier. As shown in Fig.~\ref{fig:reslight}, the fraction $I_b/I_{\text{max}}$ ranges from the $10^{-4}$ to $10^{-3}$ range, at maximum compression. This provides a lower bound as reflections from optics in the beam path likely decrease the in-ring darkness.

\section{Detection Methods and Thermometry}

\subsection{Imaging Methods and Procedure}
We use two methods of fluorescent imaging on the $X$-$A$ transition: $\Lambda$-imaging ($\Delta_{\Lambda}=25.5\,\text{MHz}$)~\cite{Cheuk2018Lambda} and resonant imaging.
For in-ring lifetime measurements with and without laser-cooling, a 100\,ms $\Lambda$-imaging pulse is used. 
For lifetime measurements in Fig.~2(a,\,c) and Fig.~\ref{fig:detuningplot}, a 50\,ms $\Lambda$-imaging pulse is used. For the normalizing image used in Fig.~2(d) and Fig.~4(f), a 20\,ms $\Lambda$-imaging pulse is used (see Section \ref{sec:ringcharacterization} for details concerning count normalization). 

We find that both methods have significant effects on the spatial distribution of molecules in free-space at the $10$'s of $\mu$m scale, with resonant imaging performing better for the same signal-to-noise. We thus primarily use resonant imaging when measuring the spatial distribution of molecules. During resonant imaging, the detuning is set to $\Delta=0\,\text{MHz}$ and all hyperfine components are present. We use a 500\,$\mu$s long imaging pulse when performing time-of-flight thermometry measurements in Fig.~2(b,\,c) and Fig.~3(a). We use a 250\,$\mu$s resonant imaging for the in-situ images shown in Fig.~1(a) and Fig.~4(b,\,c,\,d,\,e).

\begin{figure}
\includegraphics[width=0.8 \columnwidth]{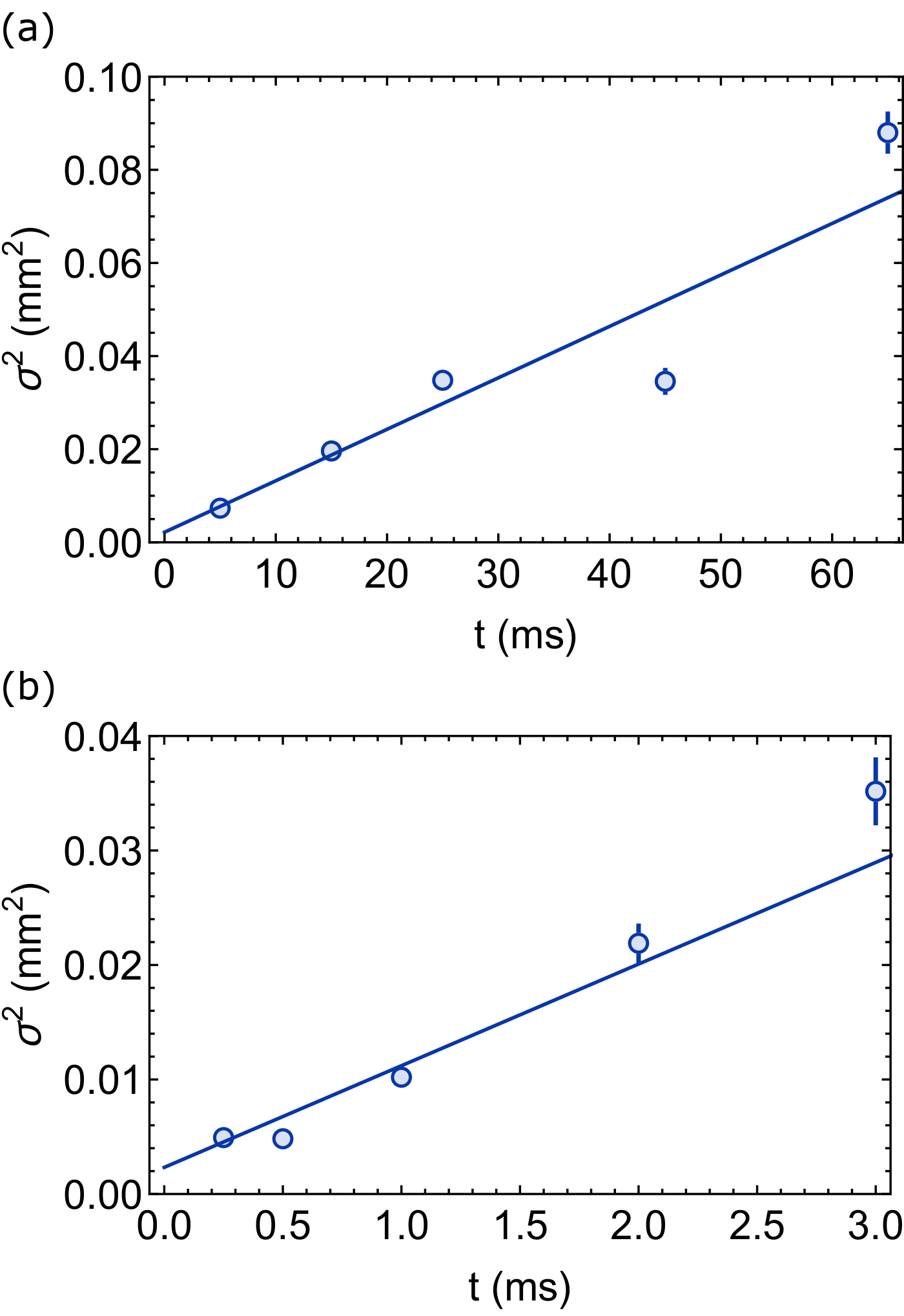}
	\caption{\label{fig:resimgblurring} Blurring under different fluorescent imaging light configurations  (see text). The extracted slope (diffusion constant ) is $1.1(2)\cdot 10^{-3}\,\text{mm}^2/\text{ms}$ for $\Lambda$-imaging (a) and   $9(1)\cdot 10^{-3}\,\text{mm}^2/\text{ms}$ for resonant imaging (b).}
\end{figure}

To quantify the effect of imaging, we release molecules from the attractive ODT and image them using the two methods for a variable time $t$. The molecules are initially loaded into the attractive ODT, after which the ODT is turned off immediately before the start of the imaging light exposure. As shown in Fig.~\ref{fig:resimgblurring}, the squared width grows linearly with time under both imaging configurations, indicative of diffusive behavior. 

Note that for Fig.~4(b,\,c,\,d), due to a large background from the repulsive barrier, the repulsive beam is kept off during resonant imaging. The observed images therefore suffer diffusive blurring, which becomes especially significant as the compression is increased and the ring becomes smaller. 

\subsection{Number Count Normalization} \label{sec:countnorm}
To reduce the effect of molecule number fluctuations, we use a normalization procedure implemented using a four-image sequence comprising two pairs of images (each with a signal and background). The signal image of the first pair records fluorescence during the initial $\Lambda$-cooling pulse following MOT loading. The signal image of the second pair is the measurement of interest. The background image of each pair is taken under identical light configuration to that of the signal image and is used for background subtraction. Hence, the first image pair can be used to monitor and correct for molecule number fluctuations. 

\subsection{Camera Count Calibration}
To extract the conversion factor between camera counts and molecule number, we measure the photon scattering rate in both the $\Lambda$-imaging and resonant imaging configurations. We first measure the lifetime of the initial $\Lambda$-cooled cloud under each configuration, but with the $v=2,\,3$ repumpers turned off. Using published values of branching ratios~\cite{pelegrini2005radiative,wall2008lifetime}, we then deduce the photon scattering rate. By combining these measurements with the detection efficiency of our camera and losses along our imaging path, we obtain a conversion factor from camera counts to molecule number for the various imaging configurations.

\subsection{Time-of-Flight Thermometry}
We measure the temperature of our molecular samples using time-of-flight expansion. We fit an integrated cut across each of the camera axes ($x',\,z$) to a 1D Gaussian curve. The squared Gaussian width $\sigma^2$ of each axis is then plotted against the effective time of flight $t_{\text{TOF}}$, which takes into account the finite imaging duration. These data are fit to a linear model $\sigma^2 = c_1 \, t_{\text{TOF}}^2 +c_0 $, and the slope $c_1$ is converted to a temperature via $T=m c_1/k_B$. The geometric mean of the extracted temperatures along the two axes $(T=\sqrt{T_{x'}T_z})$ is reported for all measurements. Note that this method is exact when both the initial spatial distribution and any blurring during imaging is described by convolution with a Gaussian kernel. These conditions are true o a good approximation for our system.

\section{Scaling Laws for Photon Scattering in a Ring Trap}
In this section, we derive the scaling laws mentioned in the main text. We assume an ideal model for a repulsive ring trap, where $U(\mathbf{r})$ is given by
\begin{eqnarray}
U(\mathbf{r}) &=&0,\,r<r_0 \nonumber \\
U(\mathbf{r}) &=& a (r-r_0)^\alpha,\,r>r_0 \nonumber
\end{eqnarray}
For a thermal ensemble, the average potential experienced by a molecule can be computed via
\begin{eqnarray}
\bra U \ket_{\text{trap}} &=& \frac{1}{Z} 2\pi \int r dr U(r) \exp\left[\frac{-U(r)}{k_B T}\right] \nonumber \\
&=& \frac{2\pi r_0^2 }{Z} \xi^{\frac{2}{\alpha}}\left[\xi^{-\frac{1}{\alpha}}\frac{\Gamma\left(2+\frac{1}{\alpha}\right)}{\alpha+1}  +\frac{\Gamma\left(2+\frac{2}{\alpha}\right)}{\alpha+2} \right] k_B T \nonumber
\end{eqnarray}
where $\xi = k_B T/ (a r_0^\alpha)$ and the partition function  $\mathcal{Z}$ is given by
\begin{eqnarray}
\mathcal{Z} &=& 2\pi \int r^2 dr U(r)\exp\left[\frac{-U(r)}{k_B T}\right] \nonumber \\
&=& 2\pi r_0^2 \left(\frac{1}{2}+ \xi^{\frac{2}{\alpha}} \left[ \xi^{-1/\alpha}\Gamma\left(\frac{1}{\alpha} +1\right)+ \frac{1}{2}\Gamma\left(1+\frac{2}{\alpha}\right)\right]\right). \nonumber 
\end{eqnarray}
In the limit of a small ring interior ($r_0 \rightarrow 0$), one finds that
\begin{equation}
\bra U \ket_{\text{trap}}  = \frac{2k_B T}{\alpha},
\end{equation}
which is independent of the overall trap potential height, $a$.

We next consider the opposite limit of a large ring interior. By large, we mean that the spatial extent in the barrier region due to finite temperature $T$ is small compared to the interior size $r_0$. In other words, the parameter $\xi = k_B T/ (a r_0^\alpha) \ll 1$. In this limit, 
\begin{equation}
\bra U \ket_{\text{trap}}  = \frac{2k_B T}{\alpha} \left[\frac{\Gamma\left(\frac{1}{\alpha}\right)}{\alpha} \xi + O(\xi^2) \right].
\end{equation}
Therefore, to lowest order, $\bra U \ket_{\text{trap}}  \propto T^{1+1/\alpha} a^{-1/\alpha}$. In the limit of a steep wall ($\alpha\rightarrow\infty$), $\bra U \ket_{\text{trap}}$ is again independent of the overall trap potential height, $a$, and is approximately linear in temperature, $T$. We note in passing that similar temperature-scaling laws have previously been derived for repulsive traps of different geometries~\cite{friedman2002dark}. 

For completeness, we also compute $\bra U^2 \ket_{\text{trap}}$. This is useful for estimating rates of two-photon processes. In a similar manner to the derivations above, we obtain
\begin{eqnarray}
\bra U^2 \ket_{\text{trap}} 
&=& \frac{2\pi r_0^2 }{Z} \xi^{\frac{2}{\alpha}}\left[\xi^{-\frac{1}{\alpha}}\frac{\Gamma\left(3+\frac{1}{\alpha}\right)}{2\alpha+1}  +\frac{\Gamma\left(3+\frac{2}{\alpha}\right)}{2\alpha+2} \right] (k_B T)^2 \nonumber
\end{eqnarray}
In the limit of a small ring interior ($r_0 \rightarrow 0$), one finds that
\begin{equation}
\bra U^2 \ket_{\text{trap}}  = \frac{2(k_B T)^2 (2+\alpha)}{\alpha},
\end{equation}
which is again independent of the overall trap potential height, $a$. In the opposite limit ($\xi \ll 1$),
\begin{equation}
\bra U^2 \ket_{\text{trap}}  = \frac{2(k_B T)^2}{\alpha} \left[\frac{\Gamma\left(2+\frac{1}{\alpha}\right)}{\alpha} \xi + O(\xi^2) \right].
\end{equation}
To lowest order,  $\bra U \ket_{\text{trap}}  \propto T^{2+1/\alpha} a^{-1/\alpha}$. In the limit of a steep wall ($\alpha \rightarrow \infty$), $\bra U \ket_{\text{trap}}$ is again independent of the overall trap potential height, $a$, but now exhibits quadratic  temperature dependence.

\section{Optical Dipole Potential and Photon Scattering in Repulsive Optical Traps}
\subsection{Analytic Expressions for Heating and Rotational Loss Rates}
In this section, we derive analytic expressions for the optical dipole potential, heating rate, and rotational cycling loss rate due to near-resonant rotational levels in the $B\,^2\Sigma^+ (v=0)$ excited state manifold. The trap detunings explored are in the +10-100\,GHz  range from the transitions to $B(v=0,N=0)$ and $B(v=0, N=2)$, which are the only two $B\,^2\Sigma^+ (v=0)$ rotational manifolds with non-vanishing dipole matrix elements with the $X\, ^2\Sigma^+ (v=0, N=1)$ ground state manifold. Since we are near-detuned from these lines, we ignore other vibrational levels ($\sim$\,THz away) and electronic levels ($\sim$ 100\,THz away). Our derivations do not take into account hyperfine structure, nominally take into account spin-rotational structure, and approximate the spin-rotational splitting to be small compared to the detuning. 

Under the rotating-wave approximation (counter-rotating term is negligible since we are near-detuned), the optical dipole potential is given by
\begin{equation}
U(r) = -\frac{\pi c^2 \Gamma_B f_{00}}{6\omega_{XB}^3}  \left[ \frac{1}{\omega_{10}-\omega}  + \frac{2}{\omega_{12}-\omega}  \right]  I(r),
\end{equation}
where $I(r)$ is the intensity, $\omega_{10}$ is the angular frequency of the $X(v=0,N=1)\rightarrow B(v=0,N=0)$ transition, $\omega_{12}$ is the angular frequency of the $X(v=0,N=1)\rightarrow B(v=0,N=2)$ transition, $\Gamma_{B}$ is the excited state decay rate of the $B$ state, and $f_{00} = 0.998$ is the vibrational Franck-Condon factor.

Due to parity and angular momentum selection rules, off-resonant photon scattering can only populate the $B (v=0, N=0)$  and  $B (v=0, N=2)$ states. While molecules excited to the $B(v=0,N=0)$ states predominantly return to the initial $X (v=0,N=1)$ manifold, molecules excited to $B (v=0, N=2)$ can decay into either $X (v=0, N=1)$ or $X (v=0 N=3)$ manifold with significant probability, the latter of which will be observed as loss. 

We can thus compute two different photon scattering rates, the rotation-preserving (cycling heating) photon scattering rate and the rotation-changing photon scattering rate. Ignoring the counter-rotating term, we find that the cycling heating rate $\Gamma_h$ for $N=1$ is
\begin{equation}
\Gamma_{h}(r) = \frac{\pi c^2 \omega^3 \Gamma_{B}^2  f_{00}}{3\omega_{XB}^6} \left(\frac{1}{(\omega_{10}-\omega)^2}  +\frac{4/5}{(\omega_{12}-\omega)^2} \right)  I(r) \nonumber
\end{equation}
and that the rotation-changing scattering rate $\Gamma_r$ for $N=1$ is
\begin{equation}
\Gamma_{r}(r) = \frac{\pi c^2 \omega^3 }{3\omega_{XB}^6} \left(\frac{2/5}{(\omega_{12}-\omega)^2} \right) \Gamma_{B}^2 f_{00} I(r). \nonumber
\end{equation}

\subsection{Loss Rate versus Trap Detuning}
As we reasoned in the main text, rotational loss, rather than heating beyond the barrier height, is the dominant loss mechanism. Since the suppression of photon-scattering in a blue-detuned trap can be understood as molecules spending less time in regions of high trap light intensity, the observed rotational loss should still be proportional to $\Gamma_r$, derived above, albeit with a suppression factor if the peak intensity $I$ is used. 

\begin{figure}
	\includegraphics[width=0.8 \columnwidth]{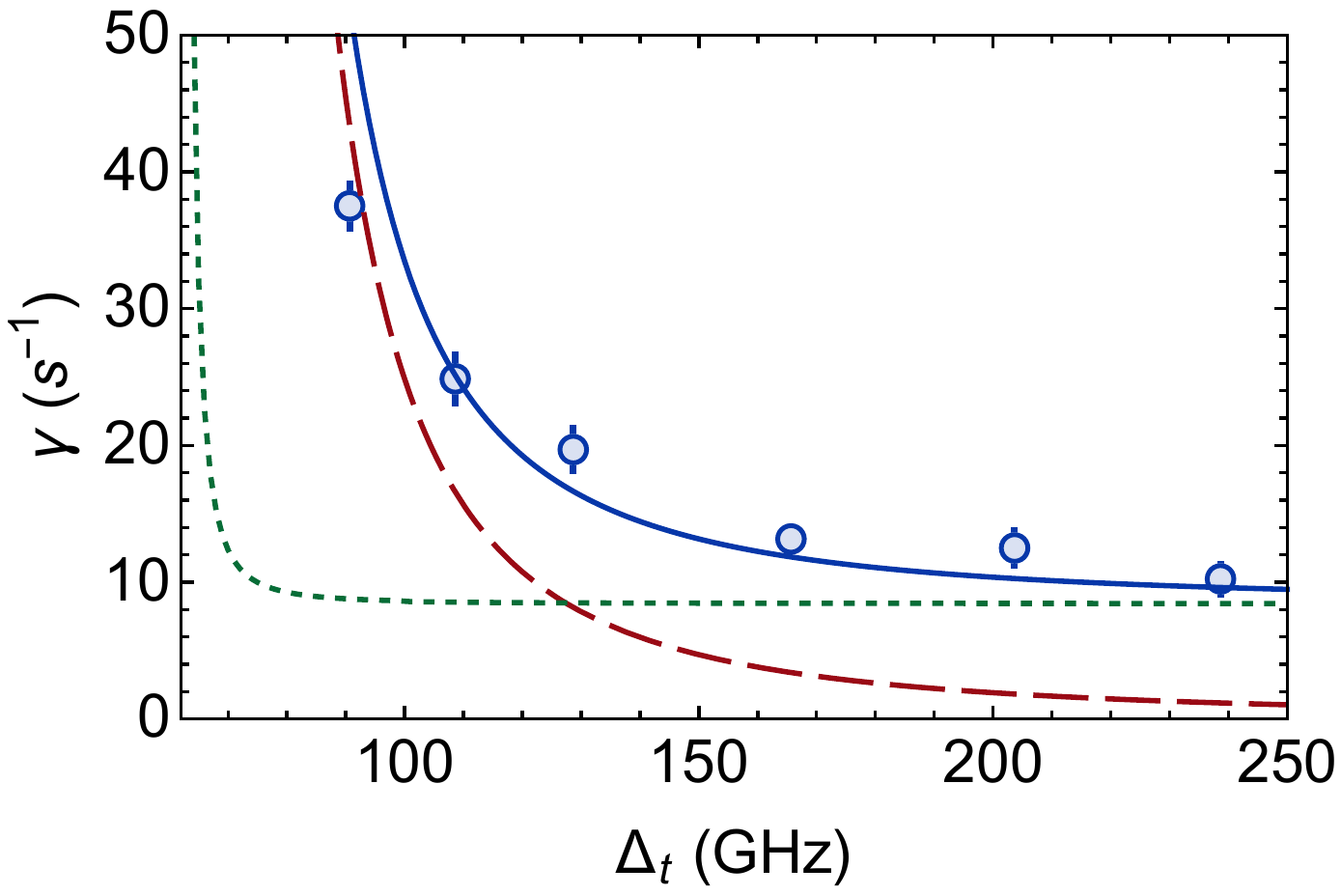}
	\caption{\label{fig:detuningplot} Measured loss rate $\gamma$ (circles) as a function of trap light detuning $\Delta_t$. The ring size is set to $r_0 = 160(10)\,\mu\text{m}$ Red dashed line shows the theoretically predicted rotational loss rate, where the effective intensity is scaled down from the peak intensity by 0.024. Green dotted line shows the expected two-photon loss due to the residual broadband spectrum of the trap laser. Here, we use a background spectral intensity of $4\times10^{-6}$ along with an effective suppression factor of $0.024^2$ (compared to peak intensity at maximum barrier height). Blue line shows the sum of the two contributions. Note that the detailed dependence depends on the spectrum of the laser, which we are unable to measure with the high resolution.}
\end{figure}

Fig. \ref{fig:detuningplot} shows the measured trap loss rates as a function of trap laser detuning $\Delta_t$. The photon loss rate is found to decay slower than the expected $1/\Delta_t^2$. We believe this is due to spectral imperfections in the trap laser, which can drive two-photon transitions that incoherently transfer molecules from $N=1$ to $N=3$. Over a bandwidth of $250\,\text{GHz}$ about the carrier, the trap laser is expected to have a decaying broadband pedestal at the $-40$ to $-60$\,dB level. As shown in Fig.~\ref{fig:detuningplot}, we can approximately reproduce the observed loss rate dependence by taking into account these incoherent two-photon 
processes. Note that the theoretical curves in Fig.~\ref{fig:detuningplot} are only for illustrative purposes, since the two-photon contribution is highly sensitive to the exact shape of the broadband pedestal, which we are unable to measure with the required resolution. In Fig.~\ref{fig:detuningplot}, the laser is assumed to have a flat broadband pedestal.

Note that the two-photon losses should scale as $T^{2+1/\alpha}\approx T^2$, as we derived above. We believe that at the detuning of $\Delta_t=108\,\text{GHz}$ used, single-photon loss dominates over two-photon loss since the measured loss rate does not reveal strong quadratic dependence on $T$.

\section{Heating Timescale versus Timescale for Rotational Loss}
During laser-cooling, spontaneous emission leads to velocity-space diffusion, which leads to heating. The velocity variance thus evolves as
\begin{equation}
\frac{d\bra v^2 \ket}{dt} = C \Gamma_h (\Delta v)^2
\end{equation}
where $\Delta v=\frac{\hbar k}{m}$ is the recoil velocity due to a resonant photon, and $C$ is a constant of order unity. It follows that the kinetic energy $K$ evolves as
\begin{equation}
\dot{K} = \frac{m}{2}\frac{d\bra v^2 \ket}{dt}= C \Gamma_h  E_R
\end{equation}
where $E_R =\frac{\hbar^2 k^2}{2m}$, $k = 2\pi/\lambda$ with $\lambda= 531\,\text{nm}$ for the $X$-$B$ transition. Hence, we expect significant heating (compared to the initial temperature $T$) to occur over a timescale 
$\frac{1}{\Gamma_h} k_B T/E_R$.
In our system,  the recoil temperature is only $E_R/k_B= 0.58\,\mu\text{K}$ while $T \approx 10\,\mu\text{K}$, implying that the timescale for significant heating is much longer than $1/\Gamma_h$. Since the rotational loss photon scattering rate and cycling photon scattering rate are comparable ($\Gamma_h \sim \Gamma_r$), we expect rotational loss to become significant much earlier than heating at laser-cooling temperatures.

\section{Laser-Cooling for Dynamical Trap Compression}
\subsection{Bounds on the Required Cooling Rate}
To estimate how laser-cooling affects dynamic trap compression, we derive both minimum and maximum bounds on the cooling rate. The problem can be modeled by a particle of mass $m$ subject to a damping acceleration of the form $a_\text{damp}=-\beta\dot{x}$, and a time-dependent force due to a moving optical barrier. The damping coefficient, $\beta$, can be recast as a cooling rate, since velocity damping leads to negative power out of the molecular sample at a rate $-\beta mv^2$.

For simplicity, we consider the situation in 1D. For concreteness, suppose the particle is positioned to the right of a barrier that is moving to the right at a speed of $v_b$. On the right slope of the barrier, the barrier provides a maximum force $F_\text{b,max}$ at its steepest point. Transforming to the frame moving with the barrier, the equation of motion at the point of maximum force reads
\begin{equation}
\ddot{x}=-\beta v_b+a_\text{b,max}.
\end{equation}
where $a_\text{b,max}= F_\text{b,max}/m$. The condition that the particle remains confined to the right, $\ddot{x}\geq 0$, sets an upper bound $\beta_{\text{max}}$ for the damping coefficient
\begin{equation}
\beta\leq\beta_{\text{max}}=\frac{a_\text{b,max}}{v_b}.
\end{equation}

We can also provide a bound for $\beta$ from below subject to the following conditions. In order to minimize barrier height requirements, we would like enough cooling such that the temperature remains constant throughout the compression. The minimum temperature increase can be found by requiring the phase space density (and hence entropy) of the sample to be constant. This corresponds to adiabatically compressing the sample. In the absence of damping, for an $n$-dimensional isotropic sample of initial radius $r_i$ and temperature $T_i$, the final temperature $T_f$ is given by
\begin{equation}
T_f=T_i\left(\frac{r_i}{r_f}\right)^2.
\end{equation}
The adiabatic rate of kinetic energy increase $\dot{K}$ is thus
\begin{equation}
\dot{K} = k_B\dot{T} = -2 k_B T \frac{\dot{r}}{r}\nonumber.
\end{equation}
Assuming the damping is large enough to counteract the adiabatic heating, the gas remains at a fixed temperature, and the maximum rate of internal energy increase is 
\begin{equation}\label{eq:enincrease}
k_B\dot{T}_{max}=-k_BT_i\,\frac{2 v_b}{r_f}.
\end{equation}
For a 2D gas, we can relate the cooling rate to the damping coefficient via
\begin{equation} \label{eq:endecrease}
k_b\dot{T}=-\beta mv^2=2\beta k_b T,
\end{equation}
where the last equality follows from the equipartition theorem. Equating the rate of internal energy increase (Eq. \ref{eq:enincrease}) and decrease (Eq. \ref{eq:endecrease}), we obtain a lower bound for the damping rate $\beta_{\text{min}}$.
\begin{equation}
\beta\geq\beta_{\text{min}} =  \frac{v_b}{r_f}.
\end{equation}

We next estimate these bounds in our system. Using the measured ring profile and the approximate ramp time, we find $\beta_{\text{min}} \approx 100\, \text{s}^{-1}$ and $\beta_{\text{max}} \sim 10^{10}\, \text{s}^{-1}$. The typical cooling timescales observed for $\Lambda$-cooling in CaF~\cite{Anderegg2018ODT} and YO~\cite{Ding2020YOSD} are on the order of  0.1\,ms, giving $\beta \sim 10^4\,\text{s}^{-1}$, comfortably within these bounds. In fact, these cooling rates are high enough such that the compression sequence used in this work could be sped up more than tenfold without significant heating. 

\subsection{Molecular Temperature During Trap Compression}
\begin{figure}
	\includegraphics[width=0.8 \columnwidth]{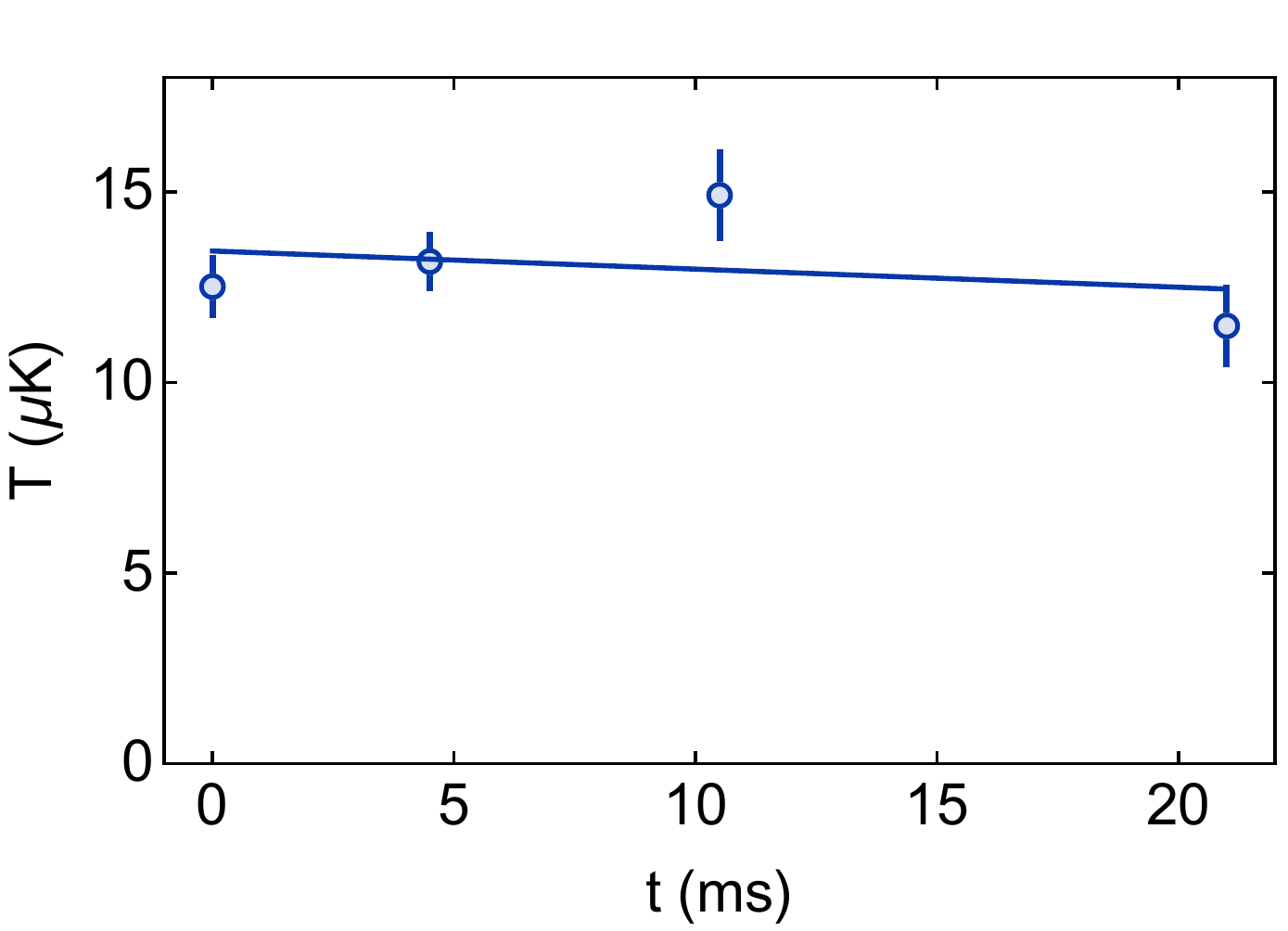}
	\caption{\label{fig:compressionT} Molecular temperature $T$ versus elapsed compression time $t$. Solid line shows a linear fit. The corresponding radii are  .}
\end{figure}

To verify that the cooling rate offered by in-trap $\Lambda$-cooling is sufficient, we measure the temperature of the molecules at various points of the compression sequence. In detail, we first load the molecules into the ring trap from the $\Lambda$-cooled cloud and immediately apply the 21\,ms compression ramp is carried out in the presence of $\Lambda$-cooling. The ramp however is truncated at various earlier times $t$. Subsequently, the cooling light is turned off, and the untrapped molecules outside the ring fall away over 20\,ms. The temperature of the remaining molecules are then measured via time-of-flight expansion. As shown in Fig.~\ref{fig:compressionT}, we observe that the temperature remains unchanged at the initial temperature of $10\,\mu \text{K}$ throughout the ramp. This verifies that indeed $\beta$ is large enough to counteract heating that arises from compression.

\bibliographystyle{apsrev4-1}